\documentstyle[12pt,citesort]{article}
\newcommand{\sotimes}{\mathop{\otimes}_{s}}
\newcommand{\cL}{\cal{L}}
\newcommand{\cR}{\cal{R}}
\newcommand{\cP}{\cal{P}}

\newcommand{\G}{\Gamma}
\begin{document}
\begin{titlepage}
\renewcommand{\thefootnote}{\fnsymbol{footnote}}
\setcounter{footnote}{1}
\vspace*{50pt}

\begin{center}
{\Large \bf SO(4) Symmetry of the Transfer Matrix for the
One-Dimensional Hubbard Model} \\
\vspace{20pt}

\footnotetext[1]{JSPS fellow}

{\Large
Masahiro {\sc Shiroishi}\footnotemark[1]%
\footnote{e-mail: {\tt siroisi@monet.phys.s.u-tokyo.ac.jp}},
Hideaki {\sc Ujino}\footnotemark[1]%
\footnote{e-mail: {\tt ujino@monet.phys.s.u-tokyo.ac.jp}}
and Miki {\sc Wadati}%
\footnote{e-mail: {\tt wadati@monet.phys.s.u-tokyo.ac.jp}}}

\vspace{20pt}

{\large\it Department of Physics, Graduate School of Science,\\
University of Tokyo, \\
Hongo 7--3--1, Bunkyo-ku, Tokyo 113, Japan}
\end{center}
\vspace{12pt}

\vfill

\begin{center}
{\bf Abstract} 
\end{center}

{\begin{quotation}

The ${SO(4)}$ invariance of the transfer matrix for 
the one-dimensional Hubbard model is clarified from the QISM (quantum 
inverse scattering method) point of view. 
We demonstrate the ${SO(4)}$ symmetry by means of the fermionic 
${L}$-operator and the fermionic ${R}$-matrix, which satisfy 
the graded Yang-Baxter relation. 
The transformation law of the fermionic ${L}$-operator 
under the ${SO(4)}$ rotation is identified with 
a kind of gauge transformation, which determines the corresponding 
transformation of the fermionic creation and annihilation operators 
under the ${SO(4)}$ rotation. 
The transfer matrix is confirmed to be invariant under the ${SO(4)}$ 
rotation, which ensures the ${SO(4)}$ invariance of the conserved 
currents including the Hamiltonian. 
Furthermore, we show that the representation of the higher 
conserved currents in terms of the Clifford algebra gives 
manifestly ${SO(4)}$ invariant forms. 

\end{quotation}}

\vspace{50pt}

\end{titlepage}

\newpage

\begin{flushleft}
{\large \bf \S 1. Introduction}
\end{flushleft}
\setcounter{equation}{0}
\renewcommand{\theequation}{1.\arabic{equation}}
\renewcommand{\thefootnote}{\fnsymbol{footnote}}
The one-dimensional (1D) Hubbard model is one of the most important
solvable models in condensed matter physics.
The ground state energy of the 1D Hubbard model
\begin{equation}
  {\cal H} = - \sum_{m=1}^{N} \sum_{s = \uparrow \downarrow} 
  (c_{ms}^{\dagger} c_{m+1s} + c_{m+1s}^{\dagger} c_{ms}) 
  + U \sum_{m=1}^{N} ( n_{m \uparrow} - \frac{1}{2} ) 
  ( n_{m \downarrow} - \frac{1}{2} ),
  \label{eq.Hubbard}
\end{equation}
with the periodic boundary condition,
\begin{equation}
  c^{\dagger}_{N+1s} = c_{1s}^{\dagger}, 
  \ \ c_{N+1s} = c_{1s} \ \ \ \ 
  (s = \uparrow \downarrow),
  \label{eq.PBC}
\end{equation}
was given by Lieb and Wu \cite{Lieb} by means of the coordinate
Bethe ansatz method.
Here ${c_{ms}^{\dagger}}$ and ${c_{ms}}$ are the fermionic creation
and annihilation operators with spin ${s (= \uparrow \downarrow)}$
at site ${m (= 1, 2, \cdots, N)}$ satisfying the canonical
anticommutation relations
\begin{equation}
  \Big\{ c_{ms}^{\dagger}, c_{m^{'} s^{'}} \Big\} = 
  \delta_{m m'} \delta_{s s'}, \ \ 
  \Big\{ c_{ms}^{\dagger}, c_{m^{'} s^{'}}^{\dagger} \Big\} = 
  \Big\{ c_{ms}, c_{m^{'} s^{'}} \Big\}  = 0,
  \label{eq.ACR}
\end{equation}
and ${n_{m s}}$ is the number density operator
\begin{equation}
  n_{ms} = c_{ms}^{\dagger} c_{ms} \ \ \ \ 
  (s = \uparrow \downarrow).
   \label{eq.density}
\end{equation}
The parameter ${U}$ is the coupling constant describing the Coulomb
interaction.

The Hamiltonian (\ref{eq.Hubbard}) enjoys 
two ${su(2)}$ symmetries 
\cite{Heilmann,Yang1,Yang2,Pernici,Affleck}. 
One is the spin-${su(2)}$ generated by
\begin{equation}
  S^{+} = \sum_{m=1}^{N} c_{m \uparrow}^{\dagger} c_{m \downarrow}, \ \ 
  S^{-} = \sum_{m=1}^{N} c_{m \downarrow}^{\dagger} c_{m \uparrow}, \ \ 
  S^{z} = \frac{1}{2} \sum_{m=1}^{N} 
  \left( n_{m \uparrow} - n_{m \downarrow} \right), 
  \label{eq.spinsu2}
\end{equation}
and the other is charge-${su(2)}$ (${\eta}$-pairing ${su(2)}$) 
generated by
\begin{equation}
  \eta^{+} = \sum_{m=1}^{N} (-1)^{m} 
  c_{m \uparrow}^{\dagger} c_{m \downarrow}^{\dagger}, \ \ 
  \eta^{-} = \sum_{m=1}^{N} (-1)^{m} c_{m \downarrow} c_{m \uparrow},  
  \ \ \eta^{z} = \frac{1}{2} \sum_{m=1}^{N} 
  \left( n_{m \uparrow} + n_{m \downarrow} - 1 \right). 
  \label{eq.etasu2}
\end{equation}
When we assume the periodic boundary condition (\ref{eq.PBC}), 
the number of sites ${N}$ should be even for
the consistency of the charge-${su(2)}$. 
In this case, the spin-${su(2)}$ and the charge-${su(2)}$ are 
connected through the partial particle-hole transformation
\begin{equation}
  c_{m \uparrow} \rightarrow c_{m \uparrow}, \ \ \ \ 
  c_{m \downarrow} \rightarrow (-1)^{m} c_{m \downarrow}^{\dagger},
  \ \ \ \ U \rightarrow -U.
  \label{eq.particlehole}
\end{equation} 
As is well known, these two ${su(2)}$ are not independent and
should be considered as elements of a bigger algebra
${so(4)}$ \cite{Yang2},
\begin{equation}
  so(4) = su(2) \oplus su(2).    
  \label{eq.so(4)}
\end{equation}
The ${so(4)}$ symmetry may be the most fundamental property of the
1D Hubbard model that characterizes the various physical features
of the model \cite{Korepin}. For example, it was proved by E{\ss}ler 
et al. \cite{Essler1,Essler2,Essler3} that
the Bethe ansatz states of the 1D Hubbard model are incomplete and
have to be complemented by the ${so(4)}$ symmetry. 
E{\ss}ler and Korepin \cite{Essler4,Essler5} showed
that the elementary excitations of the half-filled band constitute
the multiplets of ${so(4)}$.

Several authors have discussed the generalization of the Lie algebra 
symmetry ${so(4)}$ to the group symmetry ${SO(4)}$.
Following Affleck et al. \cite{Affleck2}, 
we introduce the ${2 \times 2}$ matrices
\begin{equation}
  \Psi_{2n-1} =  \left(
    \begin{array}{cc}
      c_{2n-1 \downarrow}^{\dagger} & {\rm i} c_{2n-1 \uparrow} \\
      {\rm i} c_{2n-1 \uparrow}^{\dagger}  & c_{2n-1 \downarrow}
    \end{array}
    \right), \ \ 
  \Psi_{2n} = \left(
    \begin{array}{cc}
      c_{2n \downarrow}^{\dagger} & -{\rm i} c_{2n \uparrow} \\
      {\rm i} c_{2 n \uparrow}^{\dagger} & - c_{2n \downarrow}
    \end{array}
    \right), \ \ \ \ \ n=1, \cdots, \frac{N}{2}. \label{eq.Psi}
\end{equation}
For convenience, the definition of ${\Psi_m}$ in this paper 
is chosen to be different from the usual one \cite{Affleck,Affleck2}. 
However, they are essentially equivalent.

The spin-${SU(2)}$ transformation can be realized by the left
multiplication of an ${SU(2)}$ matrix
\[
  \Psi_{m} \longrightarrow {\cal O}_{\rm spin} \Psi_{m}, \ \ \ \ 
  {\cal O}_{\rm spin} \in SU(2),
\]
while the charge-${SU(2)}$  transformation corresponds to the right
multiplication of another ${SU(2)}$ matrix,
\[
  \Psi_{m} \longrightarrow  \Psi_{m} {\cal O}_{\rm charge}, \ \ \ \ 
  {\cal O}_{\rm charge} \in SU(2).
\]
Since the left and the right matrix multiplications
are commutative, the transformation
\begin{equation}
  \tilde{\Psi}_{m} = {\cal O}_{\rm spin} \Psi_{m} {\cal O}_{\rm charge}
  \label{eq.SO(4)}
\end{equation}
gives the ${SU(2) \times SU(2)}$ transformation among the fermion 
operators. To put it more precisely, the exact group symmetry is 
\[
  SO(4) = \left[ SU(2) \times SU(2) \right] %
  / {\bf Z}_{2}, \label{eq.defSO4}
\]
because the choices ${\cal O}_{\rm spin} = -{\bf 1}$,
${\cal O}_{\rm charge}= {\bf 1}$ and 
${\cal O}_{\rm spin} = {\bf 1}$,
${\cal O}_{\rm charge}= -{\bf 1}$ 
induce the same transformation.
The infinitesimal transformation of (\ref{eq.SO(4)}) 
gives the Lie algebra symmetry (\ref{eq.so(4)}).

The integrability of the 1D Hubbard model with periodic boundary
condition was established 
by Shastry \cite{Shastry1,Shastry2,Shastry3} 
and Olmedilla et al. \cite{Olmedilla1,Olmedilla2}.
Shastry introduced a Jordan-Wigner transformation
\begin{equation}
  c_{m \uparrow} = 
  (\sigma_{1}^{z} \cdots \sigma_{m-1}^{z}) \sigma_{m}^{-},\ \ \ \ 
  c_{m \downarrow} = 
  (\sigma_{1}^{z} \cdots \sigma_{N}^{z})
  ( \tau_{1}^{z} \cdots \tau_{m-1}^{z}) \tau_{m}^{-},
  \label{eq.Jordan}
\end{equation}
which changes the fermionic Hamiltonian (\ref{eq.Hubbard}) into an 
equivalent coupled spin model
\begin{equation}
  H = \sum_{m=1}^{N} (\sigma_{m+1}^{+} \sigma_{m}^{-} + \sigma_{m}^{+}
  \sigma_{m+1}^{-}) + \sum_{m=1}^{N} (\tau_{m+1}^{+} \tau_{m}^{-} 
  + \tau_{m}^{+}\tau_{m+1}^{-}) + 
  \frac{U}{4} \sum_{m=1}^{N} \sigma_{m}^{z} \tau_{m}^{z}, 
  \label{eq.Coupled}
\end{equation}
where $\sigma$ and $\tau$ are two species of the Pauli matrices 
commuting each other, and $\sigma^{\pm}$ and $\tau^{\pm}$ are 
defined by
\[
  \sigma_{m}^{\pm} = 
  \frac{1}{2} (\sigma_{m}^{x} \pm {\rm i} \sigma_{m}^{y}), \ \ \ \
  \tau_{m}^{\pm} = \frac{1}{2} (\tau_{m}^{x} \pm {\rm i} \tau_{m}^{y}).
\]
For this equivalent coupled spin model, 
Shastry constructed the ${L}$-operator and the ${R}$-matrix 
(see Appendix), which satisfy the Yang-Baxter relation
\begin{equation}
  \check{R}_{12}(\theta_1,\theta_2) 
  \left[ L_{m}(\theta_1) \otimes L_{m}(\theta_2) \right] 
  = \left[ L_{m}(\theta_2) \otimes L_{m}(\theta_1) \right] 
  \check{R}_{12}(\theta_1,\theta_2). \label{eq.ShastryYBR}
\end{equation}
The Yang-Baxter equation for Shastry's ${R}$-matrix was recently 
proved in \cite{Shiroishi1,Shiroishi2,Shiroishi3}.

The coupled spin model (\ref{eq.Coupled}) is also referred to 
as the 1D Hubbard model, since they are related through 
the Jordan-Wigner transformation (\ref{eq.Jordan}).
However, there are differences between the coupled spin model 
(\ref{eq.Coupled}) and the fermionic Hamiltonian (\ref{eq.Hubbard}). 
It is well known that the periodic boundary condition 
for the fermion model does not correspond to the periodic boundary 
condition for the equivalent spin model. 
Moreover, due to the non-locality 
of the Jordan-Wigner transformation (\ref{eq.Jordan}), the generators 
of the ${so(4)}$ symmetry, (\ref{eq.spinsu2}) and (\ref{eq.etasu2}), 
become the non-local in terms of the spin operators 
${\sigma}$ and ${\tau}$. Thus it is more transparent to employ 
the fermionic formulation of the Yang-Baxter
relation developed by Olmedilla et al. \cite{Olmedilla1}, 
when we investigate the ${SO(4)}$ or other symmetries 
of the 1D Hubbard model from the QISM 
(quantum inverse scattering method) point of view. 
Recently, G${\ddot{\rm o}}$hmann and Murakami 
\cite{Gohmann} demonstrated 
that the transfer matrix constructed
from the fermionic ${L}$-operators has the ${su(2) \oplus su(2)}$
symmetry.
The main purpose of this paper is to generalize their result to the 
finite symmetry, namely the ${SO(4)}$ symmetry 
corresponding to (\ref{eq.SO(4)}).     

The plan of this paper is as follows. In section 2, 
we give a brief summary of the fermionic formulation 
of the QISM for the 1D Hubbard model. Some important 
properties of the fermionic ${R}$-matrix are explained.
In section 3, we prove the ${SO(4)}$ invariance of the fermionic
transfer matrix. It is shown that the ${SO(4)}$ rotation 
for the fermion operators is related to a kind of gauge 
transformation of the fermionic ${L}$-operator. 
When the number of sites is even, we can establish 
the ${SO(4)}$ symmetry of the transfer matrix 
under the periodic boundary condition.
When the number of sites is odd, we have to 
impose a twisted periodic boundary condition to establish
the ${SO(4)}$ symmetry of the transfer matrix. 
In section 4, we discuss the invariance of the transfer matrix
under the partial particle-hole transformation.
In section 5, we give a new representation of some
higher conserved currents using the Clifford algebra. 
The ${SO(4)}$ invariance of the conserved currents becomes
obvious in this representation. The last section is devoted 
to discussions.

\vspace{20pt}
\begin{flushleft}
{\large \bf \S 2. Graded Yang-Baxter Relation for the 1D Hubbard Model}
\end{flushleft}
\setcounter{equation}{0}
\renewcommand{\theequation}{2.\arabic{equation}}
As a preparation for later sections,
we shall summarize the fermionic formulation of the 1D Hubbard
model \cite{Wadati,Olmedilla1,Olmedilla2}.
The fermionic ${L}$-operator is 
\begin{equation}
  {\cL}_{m}(\theta) = 
  \left(
    \begin{array}{cccc}
      - {\rm e}^{h} f_{m \uparrow}(\theta) f_{m \downarrow}(\theta) &  
      - f_{m \uparrow}(\theta) c_{m \downarrow} &
      {\rm i} c_{m \uparrow} f_{m \downarrow}(\theta) &
      {\rm i} {\rm e}^{h} c_{m \uparrow} c_{m \downarrow} \\
      - {\rm i} f_{m \uparrow}(\theta) c_{m \downarrow}^{\dagger} &
      {\rm e}^{-h} f_{m \uparrow}(\theta) g_{m \downarrow}(\theta) &
      {\rm e}^{-h} c_{m \uparrow} c_{m \downarrow}^{\dagger} &
      {\rm i} c_{m \uparrow} g_{m \downarrow}(\theta) \\
      c_{m \uparrow}^{\dagger} f_{m \downarrow}(\theta) &
      {\rm e}^{-h} c_{m \uparrow}^{\dagger} c_{m \downarrow} &
      {\rm e}^{-h} g_{m \uparrow}(\theta) f_{m \downarrow}(\theta) &
      g_{m \uparrow}(\theta) c_{m \downarrow} \\
      - {\rm i} {\rm e}^{h} c_{m \uparrow}^{\dagger} 
      c_{m \downarrow}^{\dagger} &
      c_{m \uparrow}^{\dagger} g_{m \downarrow}(\theta) &
      {\rm i} g_{m \uparrow}(\theta) c_{m \downarrow}^{\dagger} &
      - {\rm e}^{h} g_{m \uparrow}(\theta) g_{m \downarrow}(\theta)
    \end{array}
  \right),  \label{eq.fermionicL}
\end{equation}
where
\begin{eqnarray*}
  f_{ms}(\theta) & = & \sin \theta - \left\{ \sin \theta - {\rm i} 
  \cos \theta \right\} n_{ms}, \nonumber \\
  g_{ms}(\theta) & = & \cos \theta - \left\{ \cos \theta + {\rm i} 
  \sin \theta \right\} n_{ms}. 
\end{eqnarray*}
The parameter ${h}$ is related to the spectral parameter ${\theta}$ 
and the Coulomb coupling constant ${U}$ through the relation
\begin{equation}
  \frac{\sinh 2h}{\sin 2 \theta} = \frac{U}{4}. \label{eq.constraint1}
\end{equation}
We express by ${\displaystyle \sotimes}$ the Grassmann (graded) 
direct product
\begin{eqnarray}
  & & \left[ A \otimes B \right]_{\alpha \gamma, \beta \delta} 
  = (-1)^{\left[ P(\alpha) + P(\beta) \right] P(\gamma)} 
  A_{\alpha \beta} B_{\gamma \delta}, \nonumber \\
  & & P(1) = P(4) = 0, \ \ P(2) = P(3) = 1. \label{eq.parity}
 \end{eqnarray}
The fermionic ${L}$-operator satisfies 
the graded Yang-Baxter relation \cite{Olmedilla1} 
\begin{equation}
  {\check{\cR}}_{12}(\theta_1,\theta_2) [ {\cL}_{m} (\theta_1) 
  \sotimes {\cL}_{m} (\theta_2) ] = [ {\cL}_{m} (\theta_2) 
  \sotimes {\cL}_{m} (\theta_1) ] {\check{\cR}}_{12} 
  (\theta_1,\theta_2),
  \label{eq.GYBR}
\end{equation}
under the constraint of the spectral parameter
\begin{eqnarray}
  \frac{\sinh 2 h_{1}}{\sin 2 \theta_{1}} = 
  \frac{\sinh 2 h_{2}}{\sin 2 \theta_{2}} = \frac{U}{4}.
  \label{eq.constraint2}
\end{eqnarray}

The explicit form of ${\check{\cR}_{12}(\theta_1,\theta_2)}$ 
is \cite{Olmedilla1}
{\footnotesize
\begin{eqnarray}
  & & \check{\cR}_{12}(\theta_1,\theta_2) = \nonumber \\
  & & \left( 
  \begin{array}{cccccccccccccccc}
    a^{+}&0&0&0&0&0&0&0&0&0&0&0&0&0&0&0 \\
    0&e&0&0&{\rm i}b^{-}&0&0&0&0&0&0&0&0&0&0&0 \\
    0&0&e&0&0&0&0&0&{\rm i}b^{-}&0&0&0&0&0&0&0 \\
    0&0&0&d^{+}&0&0&-{\rm i}f&0&0&{\rm i}f&0&0&-c^{+}&0&0&0 \\
    0&-{\rm i}b^{+}&0&0&e&0&0&0&0&0&0&0&0&0&0&0 \\
    0&0&0&0&0&a^{-}&0&0&0&0&0&0&0&0&0&0 \\
    0&0&0&{\rm i}f&0&0&d^{-}&0&0&-c^{-}&0&0&-{\rm i}f&0&0&0 \\
    0&0&0&0&0&0&0&e&0&0&0&0&0&-{\rm i}b^{+}&0&0 \\
    0&0&-{\rm i}b^{+}&0&0&0&0&0&e&0&0&0&0&0&0&0 \\
    0&0&0&-{\rm i}f&0&0&-c^{-}&0&0&d^{-}&0&0&{\rm i}f&0&0&0 \\
    0&0&0&0&0&0&0&0&0&0&a^{-}&0&0&0&0&0 \\
    0&0&0&0&0&0&0&0&0&0&0&e&0&0&-{\rm i}b^{+}&0 \\
    0&0&0&-c^{+}&0&0&{\rm i}f&0&0&-{\rm i}f&0&0&d^{+}&0&0&0 \\
    0&0&0&0&0&0&0&{\rm i}b^{-}&0&0&0&0&0&e&0&0 \\
    0&0&0&0&0&0&0&0&0&0&0&{\rm i}b^{-}&0&0&e&0 \\
    0&0&0&0&0&0&0&0&0&0&0&0&0&0&0&a^{+} 
  \end{array}
  \right), \nonumber \\ 
  \label{eq.Rmatrix}
\end{eqnarray}
}
where
\begin{eqnarray}
  a^{\pm} & = & \cos^2 (\theta_1 - \theta_2) 
  \left\{ 1 \pm \tanh (h_1 - h_2) 
  \frac{ \cos (\theta_1 + \theta_2)}
  {\cos (\theta_1 - \theta_2)} \right\}, \nonumber \\
  b^{\pm} & = & \sin (\theta_1 - \theta_2) 
  \cos (\theta_1 - \theta_2) 
  \left\{  1 \pm \tanh (h_1 - h_2) 
  \frac{\sin (\theta_1 + \theta_2)}
  {\sin (\theta_1 - \theta_2)} \right\} 
  \nonumber \\
  & = & \sin (\theta_1 - \theta_2) \cos (\theta_1 - \theta_2) 
  \left\{ 1 \pm \tanh (h_1 + h_2) 
  \frac{ \cos (\theta_1 + \theta_2)}
  {\cos (\theta_1 - \theta_2)} \right\}, \nonumber \\
  c^{\pm} & = & \sin^2 (\theta_1 - \theta_2) 
  \left\{ 1 \pm \tanh (h_1 + h_2) 
  \frac{\sin (\theta_1 + \theta_2)}
  {\sin (\theta_1 - \theta_2)} \right\}, \nonumber \\
  d^{\pm} & = & 1 \pm \tanh (h_1 - h_2) 
  \frac{\cos (\theta_1 - \theta_2)}{\cos (\theta_1 + \theta_2)}, 
  \nonumber \\
  & = & 1 \pm \tanh(h_1 + h_2) 
  \frac{\sin (\theta_1 - \theta_2)}{\sin (\theta_1 + \theta_2)}, 
  \nonumber  \\      
  e & = & \frac{\cos (\theta_1 - \theta_2)}{\cosh (h_1 - h_2)}, 
  \ \ \ \ 
  f =  \frac{\sin (\theta_1 - \theta_2)}{\cosh (h_1 + h_2)}. 
  \label{eq.Boltzmannweight}
\end{eqnarray} 
The second equalities for the Boltzmann weights 
${b^{\pm}}$ and ${d^{\pm}}$ are valid due to the constraints 
(\ref{eq.constraint2}). 
There are useful relations among the Boltzmann 
weights \cite{Olmedilla1}
\begin{eqnarray}
  & & d^{\pm} = a^{\pm} + c^{\pm}, \ \ 
  e^{2} = a^{+} a^{-} + b^{+} b^{-}, \ \ 
  f^{2} = b^{+} b^{-} + c^{+} c^{-}, \nonumber \\
  & & a^{+} c^{+} + a^{-} c^{-} = (b^{+})^2 + (b^{-})^2, \ \ 
  a^{+} c^{-} + a^{-} c^{+} = 2 b^{+} b^{-}, \nonumber \\
  & & a^{+} + a^{-} + c^{+} + c^{-} =2.
\end{eqnarray}

For convenience, we introduce an equivalent fermionic ${R}$-matrix 
\begin{equation}
  {\cR}_{12}(\theta_1,\theta_2) \equiv {\cP}_{12} \check{\cR}_{12}
  (\theta_1,\theta_2), \label{eq.equivalentR}
\end{equation}
where ${{\cP}_{12}}$ is the graded permutation
\begin{equation}
  {\cP}_{\alpha \gamma, \beta \delta} = (-1)^{P(\alpha) P(\gamma)} 
  \delta_{\alpha \delta} \delta_{\gamma \beta}.
\end{equation}
In terms of ${{\cal R}_{12}(\theta_1,\theta_2)}$ (\ref{eq.equivalentR}), 
the graded Yang-Baxter relation (\ref{eq.GYBR}) is cast into
\[
  {\cal R}_{12}(\theta_1,\theta_2) 
  \left( {\cal L}_{m}(\theta_{1}) \sotimes I \right) 
  \left( I \sotimes {\cal L}_{m}(\theta_{2}) \right) = 
  \left( I \sotimes {\cal L}_{m}(\theta_{2}) \right) 
  \left( {\cal L}_{m}(\theta_{1}) \sotimes I \right) 
  {\cal R}_{12}(\theta_1,\theta_2).
\]
Here ${I}$ is the ${4 \times 4}$ identity matrix.

The fundamental properties of the fermionic 
${R}$-matrix ${{\cal R}_{12}(\theta_1,\theta_2)}$ are summarized
as follows \cite{Shiroishi4}.\\
(1) Regularity (Initial condition): 
\begin{equation}
  {\cal R}_{12}(\theta_{0},\theta_{0}) = {\cal P}_{12}.
\end{equation}
(2) Graded Yang-Baxter equation:
\begin{equation}
  {\cal R}_{12}(\theta_1,\theta_2) {\cal R}_{13}(\theta_1,\theta_3) 
  {\cal R}_{23}(\theta_2,\theta_3) = {\cal R}_{23}(\theta_2,\theta_3) 
  {\cal R}_{13}(\theta_1,\theta_3) {\cal R}_{12}(\theta_1,\theta_2), 
  \label{eq.GYBE}
\end{equation}
under the constraints
\[
  \frac{\sinh 2 h_1}{\sin 2 \theta_1} 
  = \frac{\sinh 2 h_2}{\sin 2 \theta_2} 
  = \frac{\sinh 2 h_3}{\sin 2 \theta_3} = \frac{U}{4}. 
\]
(3) Unitarity:
\begin{equation}
  {\cal R}_{12}(\theta_1,\theta_2) {\cal R}_{21}(\theta_2,\theta_1) 
  = \rho (\theta_1,\theta_2) \ I , \label{eq.unitarity}
\end{equation}
where 
\[
  {\cal R}_{21}(\theta_2,\theta_1) 
  \equiv {\cal P}_{12} {\cal R}_{12}(\theta_2,\theta_1) {\cal P}_{12},
\]
and
\[
  \rho(\theta_1,\theta_2) = \cos^{2}(\theta_1 - \theta_2) 
  \left( \cos^{2}(\theta_1 - \theta_2) - \tanh^2(h_1 - h_2) 
  \cos^2(\theta_1 + \theta_2) \right).
\]

Since the non-zero elements of the ${R}$-matrix %
(\ref{eq.equivalentR}) are even with respect %
to the parity ${P(\alpha)}$ (\ref{eq.parity}), i.e.,
\[
P(\alpha) + P(\beta) + P(\alpha^{'}) + P(\beta^{'}) = 0 \ \ 
\ \ \mbox{(mod  2)} \ \ \ \ {\rm for} \ \ \ \ 
{\cR}_{\alpha \beta; \alpha^{'} \beta^{'}}(\theta_1,\theta_2) \ne 0,
\]
the graded Yang-Baxter equation (\ref{eq.GYBE}) %
can be expressed in terms of the  matrix elements %
as \cite{Kulish}
\begin{eqnarray}
& & {\cR}_{\alpha \beta; \alpha^{''} \beta^{''}}(\theta_1,\theta_2) 
{\cR}_{\alpha^{''} \gamma; \alpha^{'} \gamma^{''}}(\theta_1,\theta_3) 
{\cR}_{\beta^{''} \gamma^{''}; \beta^{'} \gamma^{'}}(\theta_2,\theta_3) 
(-)^{P(\beta^{''})(P(\alpha^{'}) + P(\alpha^{''}))} \nonumber \\
& & = {\cR}_{\beta \gamma; \beta^{''} \gamma^{''}}(\theta_2,\theta_3) 
{\cR}_{\alpha \gamma^{''}; \alpha^{''} \gamma^{'}}(\theta_1,\theta_3) 
{\cR}_{\alpha^{''} \beta^{''}; \alpha^{'} \beta^{'}}(\theta_1,\theta_2) 
(-)^{P(\beta^{''})(P(\alpha) + P(\alpha^{''}))}.
\end{eqnarray} 
Here the summentions are taken over the repeated indices.

In our previous work \cite{Shiroishi4},
we found two important relations of the fermionic ${R}$-matrix with
constant matrices ${M}$ and ${V}$.
The first relation is the symmetry of the fermionic ${R}$-matrix
\begin{equation}
  \left[ \check{\cal R}_{12}(\theta_1,\theta_2), M 
  \sotimes M \right] = 0,
  \label{eq.symmetrydef}
\end{equation}
where the general form of the matrix ${M}$ is given by
\begin{equation}
  M = \left(
  \begin{array}{cccc}
               M_{11} &   0    &    0    &   M_{14}   \\
                0     & M_{22} & M_{23}  &     0      \\
                0     & M_{32} & M_{33}  &     0      \\
               M_{41} &   0    &    0    &   M_{44}   
  \end{array}
  \right), 
  \label{eq.symmetrymatrix}
\end{equation}
with the condition 
\[
  \Delta{M} \equiv M_{11} M_{44} - M_{41} M_{14} 
  = M_{22} M_{33} - M_{23} M_{32}. \label{eq.conditions}
\]
We call the matrix $M$ symmetry matrix.
For simplicity, we assume ${\Delta{M} = 1}$ throughout the paper.

The second relation is 
\begin{equation}
  \check{\cR}_{12}(\theta_1,\theta_2;U) 
  \left[ V \sotimes V \right] 
  = \left[ V \sotimes V \right] 
  \check{\cR}_{12}(\theta_1,\theta_2;-U), 
  \label{eq.discrete}
\end{equation}
where the general form of the matrix $V$ is given by
\begin{eqnarray}
  & & V = \left(
    \begin{array}{cccc}
                0      & V_{12}  & V_{13}    &     0      \\
                V_{21} &   0     &   0       &   V_{24}   \\
                V_{31} &   0     &   0       &   V_{34}   \\
                0      & V_{42}  &  V_{43}   &     0      
    \end{array}
    \right), \nonumber \\
  & & V_{12} V_{43} - V_{13} V_{42} = 
  V_{21} V_{34} - V_{31} V_{24}. 
  \label{eq.discretematrix}
\end{eqnarray}
In the relation (\ref{eq.discrete}),  
the ${U}$-dependence of the fermionic 
${R}$-matrix is explicitly written. 
The coupling constant of the fermionic ${R}$-matrix in the 
RHS is ${- U}$, or equivalently, 
$h_{1} \rightarrow - h_1$ and $h_{2} \rightarrow - h_{2}$. 
The matrix ${V}$ is related to the partial particle-hole 
transformation (\ref{eq.particlehole}).
The constant matrices $M$ and $V$ play an important role 
in the consideration of the symmetry of the transfer matrix 
for the 1D Hubbard model (see \S 3 and \S 4). 
We remark that the symmetry matrix of Shastry's ${R}$-matrix 
is not of the form (\ref{eq.symmetrymatrix}) (see Appendix). 
This gives one of the reasons why the fermionic formulation 
employed in this paper is more appropriate for the investigation 
of the 1D Hubbard model (\ref{eq.Hubbard}).

The monodromy matrix is defined as the ordered product 
of the fermionic ${L}$-operators
\begin{equation}
  T(\theta) = \prod_{m=1}^{N \atop \longleftarrow} 
  {\cal L}_{m}(\theta) = {\cal L}_{N}(\theta) \cdots 
  {\cal L}_{1}(\theta). 
  \label{eq.monodromy}
\end{equation} 
From the (local) graded Yang-Baxter relation (\ref{eq.GYBR}), 
we have the global relation for the monodromy matrix
\begin{equation}
  {\check{\cal R}}_{12}(\theta_1,\theta_2) [ T(\theta_1) 
  \sotimes T(\theta_2) ] = [ T(\theta_2) \sotimes T(\theta_1) ] 
  {\check{\cal R}}_{12}(\theta_1,\theta_2), 
  \label{eq.globalGYBR}
\end{equation}
or equivalently
\begin{equation}
  {\cal R}_{12}(\theta_1,\theta_2) \stackrel{1}{T}
  (\theta_1) \stackrel{2}{T}(\theta_2) = 
  \stackrel{2}{T}(\theta_2) \stackrel{1}{T}(\theta_1) 
  {\cal R}_{12}(\theta_1,\theta_2), 
  \label{eq.globalGYBR2}
\end{equation}
where
\[
  \stackrel{1}{T}(\theta)  \equiv 
  {T}(\theta_1) \sotimes I, \ \ 
  \stackrel{2}{T}(\theta)  \equiv  I \sotimes {T}(\theta_2).
\]
Define the (fermionic) transfer matrix by
\begin{equation}
  {\rm str} K T(\theta) \equiv {\rm tr} 
  \left\{ \left( \sigma^{z} \otimes \sigma^{z} \right) K 
  T(\theta) \right\}, \label{eq.transfermatrix}
\end{equation}
where the constant matrix ${K}$ assumes the form 
(\ref{eq.symmetrymatrix}) and determines 
the boundary condition \cite{Shiroishi4}. 
In particular, ${K=I}$ corresponds to the periodic boundary 
condition (\ref{eq.PBC}). Then from the global graded Yang-Baxter 
relation (\ref{eq.globalGYBR}), we can prove that the transfer matrix 
(\ref{eq.transfermatrix}) constitutes a commuting family
\begin{equation}
  \left[ {\rm str} K T(\theta_{1}), {\rm str} K T(\theta_{2}) \right] = 0,
\end{equation}
which proves the integrability of the 1D Hubbard model with the (twisted) 
periodic boundary condition.

\vspace{20pt}

\begin{flushleft}
{\large \bf \S 3. SO(4) Symmetry of the Fermionic Transfer Matrix}
\end{flushleft}
\setcounter{equation}{0}
\renewcommand{\theequation}{3.\arabic{equation}}
We shall discuss the ${SO(4)}$ symmetry of the fermionic transfer matrix 
(\ref{eq.transfermatrix}). 
Let us consider the following transformation of the fermionic ${L}$-operator
\begin{equation}
  \tilde{\cL}_{m}(\theta) = {\bar{M}}^{-1} {\cL}_{m}(\theta) M, 
  \label{eq.Ltransform}
\end{equation}
where the constant matrices ${M}$ and ${\bar{M}}$ have the form of 
the symmetry matrix (\ref{eq.symmetrymatrix}), i.e.,
\begin{eqnarray}
&& \ \ M=\left(
        \begin{array}{cccc}
               M_{11} &   0    &    0    &   M_{14}   \\
                0     & M_{22} & M_{23}  &     0      \\
                0     & M_{32} & M_{33}  &     0      \\
               M_{41} &   0    &    0    &   M_{44}   
        \end{array}
      \right), \nonumber \\
&& M_{11} M_{44} - M_{41} M_{14} = M_{22} M_{33} - M_{23} M_{32} = 1, 
\label{eq.MMSO4} 
\end{eqnarray}
and
\begin{eqnarray}
&& \ \ \bar{M}=\left(
        \begin{array}{cccc}
               \bar{M}_{11} &   0    &    0    &   \bar{M}_{14}   \\
                0     & \bar{M}_{22} & \bar{M}_{23}  &     0      \\
                0     & \bar{M}_{32} & \bar{M}_{33}  &     0      \\
               \bar{M}_{41} &   0    &    0    &   \bar{M}_{44}   
        \end{array}
      \right), \ \ \nonumber \\
&& \bar{M}_{11} \bar{M}_{44} - \bar{M}_{41} \bar{M}_{14} %
= \bar{M}_{22} \bar{M}_{33} - \bar{M}_{23} \bar{M}_{32} = 1. 
\end{eqnarray}
Since the matrices ${M}$ and ${\bar{M}}$ are the symmetry matrices, 
the transformed ${L}$-operator ${\tilde{\cL}_{m}(\theta)}$ 
(\ref{eq.Ltransform}) also satisfies the graded Yang-Baxter relation 
with the ${\it same}$ fermionic ${R}$-matrix,
\begin{equation}
  {\check{\cR}}_{12}(\theta_1,\theta_2) 
  [ \tilde{\cL}_{m} (\theta_1) \sotimes \tilde{\cL}_{m} (\theta_2) ] 
  = [ \tilde{\cL}_{m} (\theta_2) \sotimes \tilde{\cL}_{m} (\theta_1) ] 
  {\check{\cR}}_{12} (\theta_1,\theta_2). 
  \label{eq.tildeGYBR}
\end{equation}
We now look for a special transformation of (\ref{eq.Ltransform}), 
which satisfies
\begin{equation}
  {\tilde{\cL}}_{m}(\theta;c_{ms}) = {\cL}_{m}(\theta;\tilde{c}_{ms}). 
  \label{eq.Lcondition}
\end{equation}
Here we explicitly write the dependence of the fermionic 
${L}$-operator on the fermion operators.  
The fermion operators ${c_{ms}}$ and ${\tilde{c}_{ms}}$ 
are assumed to be connected through the transformation law 
(\ref{eq.SO(4)}).
We discovered that the relation 
(\ref{eq.Lcondition}) is satisfied when the matrices
${M}$ and ${\bar{M}}$ meet the following conditions
\begin{eqnarray}
  & & M_{44} = M_{11}^{*}, \ \ M_{41} = - M_{14}^{*}, \ \ 
  M_{33} = M_{22}^{*}, \ \ M_{32} = -M_{23}^{*} \nonumber \\
  & & \bar{M}_{11} = M_{11}, \ \ \bar{M}_{44} = M_{44}, \ \ 
  \bar{M}_{14} = - M_{14}, \ \ \bar{M}_{41} = - M_{41}, \nonumber \\
  & & \bar{M}_{22} = M_{22}, \ \ \bar{M}_{33} = M_{33}, \ \ 
  \bar{M}_{23} = M_{23}, \ \ \bar{M}_{32} = M_{32}. \label{eq.specialcond}
\end{eqnarray}
The condition (\ref{eq.MMSO4}) now becomes 
\begin{equation}
|M_{11}|^2 + |M_{14}|^2 = |M_{22}|^2 + |M_{23}|^2 = 1. \label{eq.unit}
\end{equation}

It is useful to introduce the submatrices of the matrices 
${M}$ and ${\bar{M}}$ as 
\begin{eqnarray}
  & & M_{\rm charge} = \left( \begin{array}{cc}
                       M_{11} & M_{14} \\
                       M_{41} & M_{44} \\
                    \end{array}
             \right), \ \ \ \ 
  M_{\rm spin} = \left( \begin{array}{cc}
                       M_{22} & M_{23} \\
                       M_{32} & M_{33} \\
                  \end{array}
           \right), \nonumber \\
  & & \bar{M}_{\rm charge} = \left( \begin{array}{cc}
                       \bar{M}_{11} & \bar{M}_{14} \\
                       \bar{M}_{41} & \bar{M}_{44} 
                    \end{array}
             \right), \ \ \ \ 
  \bar{M}_{\rm spin} = \left( \begin{array}{cc}
                       \bar{M}_{22} & \bar{M}_{23} \\
                       \bar{M}_{32} & \bar{M}_{33} 
                  \end{array}
           \right). \label{eq.submatrix}
\end{eqnarray}
Then the conditions (\ref{eq.specialcond}) and (\ref{eq.unit}) are 
equivalent to the relations  
\begin{equation}
  \bar{M}_{\rm charge} = \sigma^{z} M_{\rm charge} \sigma^{z}, 
  \ \ \ \ \bar{M}_{\rm spin} = M_{\rm spin}, \ \ \ \ \ \ \ \ \ \ 
  M_{\rm charge}, \ M_{\rm spin} \in SU(2).  
  \label{eq.MSO4}
\end{equation}
The corresponding transformation law of the fermion operators 
is 
\begin{equation}
  \left( \begin{array}{cc}
       \tilde{c}_{m \downarrow}^{\dagger} & {\rm i} \tilde{c}_{m \uparrow} \\ 
     {\rm i} \tilde{c}_{m \uparrow}^{\dagger} & \tilde{c}_{m \downarrow}  
       \end{array}
  \right)
  = \left( \begin{array}{cc}
             M_{22}^{*} & -M_{23} \\
             M_{23}^{*} &  M_{22}
             \end{array}
       \right) 
  \left( \begin{array}{cc}
       c_{m \downarrow}^{\dagger} &  {\rm i} c_{m \uparrow} \\ 
     {\rm i} c_{m \uparrow}^{\dagger} & c_{m \downarrow}  
       \end{array}
  \right)
  \left( \begin{array}{cc}
                       M_{11} & M_{14} \\
                      -M_{14}^{*} & M_{11}^{*} 
                  \end{array}
  \right). 
  \label{eq.firstfermion}
\end{equation}
Hereafter, we implicitly assume the the conditions (\ref{eq.MSO4}) 
for the matrices ${M}$ and ${\bar{M}}$.
Then the transformation (\ref{eq.Ltransform}) is not 
a gauge transformation in a strict sense, 
because
${\bar{M} \ne M}$ 
(particularly ${\bar{M}_{\rm charge} \ne M_{\rm charge}}$). 
So we try assigning the different transformation laws to
the ${L}$-operators for odd sites and even sites as
\begin{equation}
  \tilde{\cL}_{2n-1}(\theta) = \bar{M}^{-1} {\cL}_{2n-1}(\theta) M, 
  \ \ \ \ \tilde{\cL}_{2n}(\theta) = M^{-1} {\cL}_{2n}(\theta) \bar{M}. 
  \label{eq.evenoddSO4}
\end{equation}
The corresponding transformation law of the fermion operators on 
odd sites is, of course, given by the formula (\ref{eq.firstfermion}),
\begin{equation}
  \left( \begin{array}{cc}
    \tilde{c}_{2n-1 \downarrow}^{\dagger} & 
    {\rm i} \tilde{c}_{2n-1 \uparrow} \\ 
    {\rm i} \tilde{c}_{2n-1 \uparrow}^{\dagger} & 
    \tilde{c}_{2n-1 \downarrow}  
  \end{array}
  \right)
  = \left( \begin{array}{cc}
             M_{22}^{*} & -M_{23} \\
             M_{23}^{*} &  M_{22}
             \end{array}
  \right) 
  \left( \begin{array}{cc}
       c_{2n-1 \downarrow}^{\dagger} & {\rm i} c_{2n-1 \uparrow} \\ 
       {\rm i} c_{2n-1 \uparrow}^{\dagger} & c_{2n-1 \downarrow}  
       \end{array}
  \right)
  \left( \begin{array}{cc}
                       M_{11} & M_{14} \\
                      -M_{14}^{*} & M_{11}^{*} 
                  \end{array}
  \right). \label{eq.oddfermion}
\end{equation}
Since the matrices ${M}$ and ${\bar{M}}$ are related 
by the exchange ${M_{14} \leftrightarrow - M_{14}}$, 
the transformation law for even sites reads
\[
  \left( \begin{array}{cc}
    \tilde{c}_{2n \downarrow}^{\dagger} & {\rm i} 
    \tilde{c}_{2n \uparrow} \\ 
    {\rm i} \tilde{c}_{2n \uparrow}^{\dagger} & 
    \tilde{c}_{2n \downarrow}  
  \end{array}
  \right)
  = \left( \begin{array}{cc}
             M_{22}^{*} & -M_{23} \\
             M_{23}^{*} &  M_{22}
             \end{array}
    \right) 
  \left( \begin{array}{cc}
       c_{2n \downarrow}^{\dagger} & {\rm i} c_{2n \uparrow} \\ 
       {\rm i} c_{2n \uparrow}^{\dagger} & c_{2n \downarrow}  
  \end{array}
  \right)
  \left( \begin{array}{cc}
    M_{11} & -M_{14} \\
    M_{14}^{*} & M_{11}^{*} 
  \end{array}
  \right). 
  \label{eq.evenfermion}
\]
Multiplying the Pauli matrix $\sigma^{z}$ from the right, 
we obtain
\begin{equation}
  \left( \begin{array}{cc}
    \tilde{c}_{2n \downarrow}^{\dagger} &  
    - {\rm i} \tilde{c}_{2n \uparrow} \\ 
    {\rm i} \tilde{c}_{2n \uparrow}^{\dagger} & 
    -\tilde{c}_{2n \downarrow}  
  \end{array}
  \right)
  = \left( \begin{array}{cc}
             M_{22}^{*} & -M_{23} \\
             M_{23}^{*} &  M_{22}
             \end{array}
  \right) 
  \left( \begin{array}{cc}
    c_{2n \downarrow}^{\dagger} &  - {\rm i} c_{2n \uparrow} \\ 
    {\rm i} c_{2n \uparrow}^{\dagger} & -c_{2n \downarrow}  
  \end{array}
  \right)
  \left( \begin{array}{cc}
                       M_{11} & M_{14} \\
                      -M_{14}^{*} & M_{11}^{*} 
  \end{array}
  \right). 
  \label{eq.evenfermion2}
\end{equation}
Recalling the definition of the 
${2 \times 2}$ matrices ${\Psi_{m}}$ (\ref{eq.Psi}),
we can summarize the transformation laws (\ref{eq.oddfermion}) 
and (\ref{eq.evenfermion2}) as
\begin{equation}
  \tilde{\Psi}_{m} = M_{\rm spin}^{-1} \Psi_{m} M_{\rm charge}, 
  \label{eq.summaryfermion}
\end{equation}
which exactly coincides (\ref{eq.SO(4)}) with the correspondences 
${{\cal O}_{\rm spin} = M_{\rm spin}^{-1}}$ 
and ${{\cal O}_{\rm charge} = M_{\rm charge}}$. 
As we have explained in \S 1, the transformation (\ref{eq.summaryfermion}) 
is the ${SO(4)}$ rotation in the space of 
the fermion operators. Therefore we can conclude that a kind of gauge 
transformation (\ref{eq.evenoddSO4}) induces the ${SO(4)}$ 
rotations for the fermion operators. 
We call the transformation (\ref{eq.evenoddSO4}) 
the ${SO(4)}$ rotation for the fermionic ${L}$-operator 
(\ref{eq.fermionicL}). Note that the canonical anticommutation relation 
(\ref{eq.ACR}) is preserved under the transformation (\ref{eq.summaryfermion})
\[
  \Big\{ \tilde{c}_{ms}^{\dagger}, 
  \tilde{c}_{m^{'} s^{'}} \Big\} 
  = \delta_{m m^{'}} \delta_{s s^{'}}, \ \ \ \ 
  \Big\{ \tilde{c}_{ms}^{\dagger}, 
  \tilde{c}_{m^{'} s^{'}}^{\dagger} \Big\} 
  = \Big\{ \tilde{c}_{ms}, \tilde{c}_{m^{'} s^{'}} \Big\} = 0. 
  \label{eq.trACR}
\]

Let us consider the ${SO(4)}$ invariance of the fermionic 
transfer matrix (\ref{eq.transfermatrix}).  
First we assume that ${N}$ is even and impose the periodic boundary condition.
The local ${SO(4)}$ rotation for the fermionic 
${L}$-operators (\ref{eq.evenoddSO4}) induces the ${SO(4)}$ rotation 
for the monodromy matrix (\ref{eq.monodromy})
\begin{equation}
  \tilde{T}(\theta) \equiv \prod_{m=1}^{N \atop \longleftarrow} 
  \tilde{\cL}_{m}(\theta) = M^{-1} T(\theta) M. 
  \label{eq.SO4monodromy}
\end{equation} 
Since the relation
\begin{equation}
  {\rm str} \left\{ X(\theta) M \right\} 
  = {\rm str} \left\{ M X(\theta) \right\} 
  \label{eq.strM}
\end{equation}
holds, the transfer matrix (\ref{eq.transfermatrix}) is invariant 
under the periodic boundary condition (${K = I}$)
\begin{equation}
  {\rm str} \tilde{T}(\theta;c_{ms}) = {\rm str} T(\theta;c_{ms}), 
  \label{eq.monodromy2}
\end{equation}
where we write the fermion operators explicitly. 
In the relation (\ref{eq.strM}), 
${X(\theta)}$ is any 4 ${\times}$ 4 matrix, 
which may depend on the fermion operators. 
On the other hand, the transfer matrix ${{\rm str} \tilde{T}(\theta;c_{ms})}$ 
can be expressed as
\begin{equation}
  {\rm str} \tilde{T}(\theta;c_{ms}) = {\rm str} T(\theta;\tilde{c}_{ms}),
  \label{eq.monodromy3}
\end{equation}
due to the property (\ref{eq.Lcondition}).
Combining (\ref{eq.monodromy2}) and (\ref{eq.monodromy3}), 
we establish
\begin{equation}
  {\rm str} T(\theta;c_{ms}) = {\rm str} T(\theta;\tilde{c}_{ms}). 
  \label{eq.monodromy4}
\end{equation}
The relation (\ref{eq.monodromy4}) shows that the fermionic 
transfer matrix is invariant under 
the ${SO(4)}$ rotation for the fermion operators 
(\ref{eq.summaryfermion}). 
It indicates that all the higher conserved currents, 
which are embedded in the transfer matrix, 
also have the ${SO(4)}$ symmetry (see \S 5). 

Now we shall write the transformation (\ref{eq.SO4monodromy}) 
in terms of the submatrices ${M_{\rm charge}}$ and 
${M_{\rm spin}}$ (\ref{eq.submatrix}). 
We introduce the following convenient notation 
for the monodromy matrix \cite{Gohmann}
\[
  T(\theta) =  \left(
    \begin{array}{cccc}
      D_{11}(\theta) & C_{11}(\theta) & 
      C_{12}(\theta) & D_{12}(\theta) \\
      B_{11}(\theta) & A_{11}(\theta) & 
      A_{12}(\theta) & B_{12}(\theta) \\
      B_{21}(\theta) & A_{21}(\theta) & 
      A_{22}(\theta) & B_{22}(\theta) \\
      D_{21}(\theta) & C_{21}(\theta) & 
      C_{22}(\theta) & D_{22}(\theta)   
    \end{array}
    \right),
\]
where we regard
$A(\theta)= (A_{ij}(\theta))$, $B(\theta) = (B_{ij}(\theta))$,
$C(\theta) = (C_{ij}(\theta))$ and $D(\theta)=(D_{ij}(\theta))$
as ${2 \times 2}$ matrices.
Then the transformation (\ref{eq.SO4monodromy}) can be expressed 
in terms of ${2 \times 2}$ matrices ${A(\theta),\cdots,D(\theta)}$ as
\begin{eqnarray}
  \tilde{A}(\theta) &=&
  M_{\rm spin}^{-1} A(\theta) M_{\rm spin}, \nonumber \\ 
  \tilde{B}(\theta) &=&  
  M_{\rm charge}^{-1} B(\theta) M_{\rm spin}, \nonumber \\ 
  \tilde{C}(\theta) &=& M_{\rm spin}^{-1} 
  C(\theta) M_{\rm charge}, \nonumber \\ 
  \tilde{D}(\theta) &=& M_{\rm charge}^{-1} D(\theta) M_{\rm charge}.
  \label{eq.SO4element}
\end{eqnarray}
Because the transformed monodromy matrix also satisfies 
the graded Yang-Baxter relation with the fermionic ${R}$-matrix,
\[
  \check{R}_{12}(\theta_1,\theta_2) 
  \left[ \tilde{T}(\theta_1) \sotimes \tilde{T}(\theta_2) \right] 
  = \left[ \tilde{T}(\theta_2) \sotimes \tilde{T}(\theta_1) \right] 
  \check{R}_{12}(\theta_1,\theta_2),
\]
the associative algebra defined by the graded Yang-Baxter relation 
should be invariant under the transformation (\ref{eq.SO4element}).  
From (\ref{eq.SO4element}), we notice an interesting fact that 
the submatrix ${A(\theta)}$ is transformed by the spin-${SU(2)}$ 
rotation and ${D(\theta)}$ is transformed by the charge-${SU(2)}$ 
rotation. 
We believe this property plays a significant role 
in the application of the algebraic Bethe ansatz 
for the 1D Hubbard model \cite{Ramos}. 

Next we consider the case of ${N}$ odd. 
The monodromy matrix transforms as
\begin{equation}
  \tilde{T}(\theta) = \bar{M}^{-1} T(\theta) M.
\end{equation}
In this case, we have to twist the periodic boundary condition
to make the transfer matrix ${SO(4)}$ invariant. 
The condition for the matrix ${K}$ in the transfer matrix is 
\begin{equation}
  K \bar{M} = M K. \label{eq.Kcond}
\end{equation}
For example,
\begin{equation}
  K = \left( 
    \begin{array}{cccc}
      {\rm i} & 0 & 0 &  0      \\
      0   & 1 & 0 &  0      \\
      0   & 0 & 1 &  0      \\
      0   & 0 & 0 & -{\rm i} 
    \end{array}
  \right) \label{eq.Kchoice}
\end{equation}
solves the condition (\ref{eq.Kcond}). 
From  the formula in \cite{Shiroishi4}, 
we can see that the choice (\ref{eq.Kchoice}) corresponds to 
the twisted boundary condition
\begin{eqnarray}
  & & c_{N+1 \uparrow}^{\dagger} = 
  {\rm i} c_{1 \uparrow}^{\dagger}, \ \ 
  c_{N+1 \uparrow} = - {\rm i} c_{1 \uparrow}, 
  \nonumber \\
  & & c_{N+1 \downarrow}^{\dagger} = 
  {\rm i} c_{1 \downarrow}^{\dagger}, \ \ 
  c_{N+1 \downarrow} = - {\rm i} c_{1 \downarrow}. 
  \label{eq.twistedPBC}
\end{eqnarray}
Assuming (\ref{eq.Kchoice}) and (\ref{eq.twistedPBC}), 
we can prove the ${SO(4)}$ invariance of the transfer matrix 
for $N$ odd as 
\begin{equation}
  {\rm str} K T(\theta;c_{ms}) = {\rm str} K T(\theta;\tilde{c}_{ms}),
\end{equation}
in a similar way to the even case.

\vspace{20pt}

\begin{flushleft}
{\large \bf \S 4. Partial Particle-Hole Transformation 
of the Fermionic Transfer Matrix}
\end{flushleft}
\setcounter{equation}{0}
\renewcommand{\theequation}{4.\arabic{equation}}
In \cite{Gohmann}, the transformation law of 
the fermionic ${L}$-operator corresponding to
 the partial particle-hole transformation 
 (\ref{eq.particlehole}) was found. 
We shall discuss the transformation law 
in connection with the relation (\ref{eq.discrete}). 
Consider the following transformations 
of the fermionic ${L}$-operators
\begin{equation}
 \hat{\cL}_{2n-1}(\theta) = \bar{V}^{-1} {\cL}_{2n-1}(\theta) V, 
 \ \ \ \ \hat{\cL}_{2n}(\theta) = V^{-1} {\cL}_{2n}(\theta) \bar{V}, 
 \label{eq.phtransL}
\end{equation}
where 
\begin{equation}
  V = \left( 
  \begin{array}{cccc}
    0  & -1 &  0 &  0      \\
    1  &  0 &  0 &  0      \\
    0  &  0 &  0 &  1      \\
    0  &  0 & -1 &  0 
  \end{array}
  \right), \ \ \ \ 
  \bar{V} = \left( 
  \begin{array}{cccc}
    0  &  1 &  0 &  0      \\
    1  &  0 &  0 &  0      \\
    0  &  0 &  0 & -1      \\
    0  &  0 & -1 &  0 
  \end{array}
  \right).
\end{equation}
Since the constant matrices ${V}$ and ${\bar{V}}$ are of the form 
(\ref{eq.discretematrix}), we have the following graded 
Yang-Baxter relation with the transformed ${L}$-operator 
(\ref{eq.phtransL})
\begin{equation}
  {\check{\cR}}_{12}(\theta_1,\theta_2;-U) 
  [ \hat{\cL}_{m} (\theta_1;U) \sotimes 
  \hat{\cL}_{m} (\theta_2;U) ] 
  = [ \hat{\cL}_{m} (\theta_2;U) \sotimes 
  \hat{\cL}_{m} (\theta_1;U) ] 
  {\check{\cR}}_{12} (\theta_1,\theta_2;-U), 
  \label{eq.particleholeGYBR}
\end{equation}
where we write the ${U}$-dependence explicitly. 
The graded Yang-Baxter relation (\ref{eq.particleholeGYBR}) 
implies that the transformed ${L}$-operators 
${\hat{\cL}_{m}(\theta;U)}$ are related to the ${L}$-operators 
with the coupling constant ${-U}$. 
In fact the following relations hold,
\begin{eqnarray}
  \hat{\cL}_{2n-1}(\theta;c_{2n-1 s}, U) 
  &=& {\rm i} {\cL}_{2n-1}(\theta;\hat{c}_{2n-1 s},-U), \nonumber \\
  \hat{\cL}_{2n}(\theta;c_{2n s},U) 
  &=& {\rm i} {\cL}_{2n}(\theta;\hat{c}_{2n s},-U), \label{eq.phtransL2}
\end{eqnarray}
where
\begin{eqnarray}
  \hat{c}_{2n-1 \uparrow} & = & c_{2n-1 \uparrow}, \ \ 
  \hat{c}_{2n-1 \downarrow} = - c_{2n-1 \downarrow}^{\dagger}, 
  \nonumber \\
  \hat{c}_{2n \uparrow} & = & c_{2n \uparrow}, \ \ \ \ \ \ 
  \hat{c}_{2n \downarrow} = c_{2n \downarrow}^{\dagger}. 
  \label{eq.phfinal}
\end{eqnarray}
The transformation (\ref{eq.phfinal}) is nothing 
but the partial particle-hole transformation (\ref{eq.particlehole}). 
Therefore we call (\ref{eq.phtransL}) the partial particle-hole
transformation of the fermionic ${L}$-operator (\ref{eq.fermionicL}).

It is quite interesting to note that the transformation (\ref{eq.phfinal}) 
can be written in terms of the ${2 \times 2}$ matrix 
${\Psi_{m}}$ (\ref{eq.Psi}) as
\begin{equation}
\hat{\Psi}_{m} = - \Psi_{m}^{\dagger},  \label{eq.phPsi}
\end{equation}
where ${\dagger}$ denotes the hermitian conjugation.
Moreover, taking the hermitian conjugation of (\ref{eq.summaryfermion}),
we find
\begin{equation}
\tilde{\hat{\Psi}}_{m} = M_{\rm charge}^{-1} \hat{\Psi}_{m} M_{\rm spin},
\end{equation}
which shows that the spin-${SU(2)}$ ${M_{\rm spin}}$ 
and the charge-${SU(2)}$ ${M_{\rm charge}}$ are exchanged 
after the partial particle-hole transformation.  

We are ready to verify the invariance of the transfer matrix
of the 1D Hubbard model under the partial particle-hole 
transformation. First we assume that ${N}$ is even and 
impose the periodic boundary condition. 
Then the partial particle-hole transformation of the monodromy
matrix induced by (\ref{eq.phtransL}) is
\begin{equation}
\hat{T}(\theta;c_{ms},U) = V^{-1} T(\theta;c_{ms},U) V.
\end{equation}
From the relations
\[
  {\rm str} \hat{T}(\theta;c_{ms},U) 
  = {\rm str} \left\{ V^{-1} T(\theta;c_{ms},U) V \right\} 
  = - {\rm str} T(\theta;c_{ms},U),  \label{eq.phderivation}
\]
and
\[
  {\rm str} \hat{T}(\theta;c_{ms},U) 
  = {\rm i}^{N} {\rm str} T(\theta;\hat{c}_{ms},-U),
\]
we obtain \cite{Gohmann}
\begin{equation}
  {\rm str} T(\theta;c_{ms},U) 
  = - {\rm i}^{N} {\rm str} T(\theta;\hat{c}_{ms},-U). 
  \label{eq.phtransfermatrix}
\end{equation}
The last identity (\ref{eq.phtransfermatrix}) proves 
the invariance of the fermionic transfer matrix (up to sign) 
under the the partial particle-hole transformation 
(\ref{eq.particlehole}). 
Note a relation
\begin{equation}
  {\rm str} \left\{ X(\theta) V \right\} 
  = -{\rm str} \left\{ V X(\theta) \right\}, 
  \label{eq.strV}
\end{equation}
which should be compared with (\ref{eq.strM}).

We have a similar relation for ${N}$ odd,
\begin{equation}
  {\rm str} \left\{ K T(\theta;c_{ms},U) \right\} 
  = - {\rm i}^{N-1} {\rm str} \left\{ K T(\theta;\hat{c}_{ms},-U) \right\}. 
  \label{eq.phtransfermatrix2}
\end{equation}
where ${K}$ is given by (\ref{eq.Kchoice}). 
In the derivation of (\ref{eq.phtransfermatrix2}), we have used the relation
\[
K \bar{V}^{-1} = {\rm i} V^{-1} K.
\]
The formula (\ref{eq.phtransfermatrix2})  means that 
the fermionic transfer matrix for ${N}$ odd 
is also invariant under the partial particle-hole transformation 
(\ref{eq.particlehole}) when we assume the twisted periodic 
boundary condition (\ref{eq.twistedPBC}).

\vspace{20pt} 

\begin{flushleft}
{\large \bf \S 5. SO(4) Symmetry of the Higher Conserved Currents}
\end{flushleft}
\setcounter{equation}{0}
\renewcommand{\theequation}{5.\arabic{equation}}
In \S 3, we have shown the ${SO(4)}$ symmetry of the transfer matrix, 
which means that all the conserved currents of the 1D Hubbard model 
also have the ${SO(4)}$ symmetry. 
As will be seen, the ${SO(4)}$ symmetry of the conserved currents 
can be manifestly read out in terms of the Clifford algebra.
Hereafter, for simplicity of explanation,
we assume the number of sites is always even 
and impose the periodic boundary condition.

Although the graded Yang-Baxter relation ensures the existence of
infinitely many higher conserved currents in involution, 
it is not an easy task to obtain their explicit forms from the 
transfer matrix (see \cite{Olmedilla2}). 
To construct the higher conserved currents, 
we often use the boost operator \cite{Tetelman,Sogo},
which recursively produces the higher conserved currents.
However in the case of the 1D Hubbard model,
the boost operator does not exist \cite{Grabowski} 
and we have to resort to a more direct computation.

The first higher conserved current of the 1D Hubbard model 
was found by Shastry \cite{Shastry1,Shastry3} as
\begin{eqnarray}
  I^{(2) }& = & {\rm i} \sum_{m=1}^{N} 
  \sum_{s=\uparrow \downarrow} \left( 
  c_{m+2 s}^{\dagger} c_{m s} - c_{m s}^{\dagger} c_{m+2 s} 
  \right) \nonumber \\
  & & - {\rm i} U \sum_{m=1}^{N} \sum_{s=\uparrow \downarrow} 
  \left( c_{m+1 s}^{\dagger} c_{m s} 
  - c_{m s}^{\dagger} c_{m+1 s} \right) 
  \left( n_{m+1,-s} + n_{m,-s} - 1 \right). 
  \label{eq.I2}
\end{eqnarray}
Subsequently, some higher conserved currents 
were obtained in a similar fashion \cite{Grabowski,Grosse,Zhou},
\begin{eqnarray}
  I^{(3)} & = & - \sum_{m=1}^{N} \sum_{s=\uparrow \downarrow} 
  \left( c_{m+3 s}^{\dagger} c_{m s} 
  + c_{m s}^{\dagger} c_{m+3 s} \right) \nonumber \\
  & & + U \sum_{m=1}^{N} \sum_{s=\uparrow \downarrow} 
  \Bigg\{ \left( c_{m+1 s}^{\dagger} c_{m-1 s} 
  + c_{m-1 s}^{\dagger} c_{m+1 s} \right) 
  \left( n_{m+1,-s} + n_{m,-s} + n_{m-1,-s} 
  - \frac{3}{2} \right) \nonumber \\
  & & \ \ \ \ \ \ \ \ \ \ \ \ \ \ \ \ 
  + \left( c_{m+1 s}^{\dagger} c_{m s} - c_{m s}^{\dagger} c_{m+1 s} 
  \right) \left( c_{m,-s}^{\dagger} c_{m-1,-s} - c_{m-1,-s}^{\dagger} 
  c_{m,-s} \right) \nonumber \\ 
  & & \ \ \ \ \ \ \ \ \ \ \ \ \ \ \ \ 
  - \left( n_{m+1 s} - \frac{1}{2} \right) 
  \left( n_{m, -s} - \frac{1}{2} \right) \Bigg\} 
  \nonumber \\
  & & + U \sum_{m=1}^{N} 
  \Bigg\{ \left( c_{m+1 \uparrow}^{\dagger} 
  c_{m \uparrow} - c_{m \uparrow}^{\dagger} c_{m+1 \uparrow} 
  \right) \left( c_{m+1 \downarrow}^{\dagger} c_{m \downarrow} 
  - c_{m \downarrow}^{\dagger} c_{m+1 \downarrow} \right)\nonumber\\
  & & \hspace*{40pt} 
  - \left( n_{m \uparrow} - \frac{1}{2} \right) 
  \left( n_{m \downarrow} - \frac{1}{2} \right) \Bigg\} 
  \nonumber \\
  & & - U^{2} \sum_{m=1}^{N} \sum_{s= \uparrow \downarrow} 
  \left( c_{m+1 s}^{\dagger} c_{m s} 
  + c_{ms}^{\dagger} c_{m+1 s} \right) 
  \left( n_{m,-s} - \frac{1}{2} \right) 
  \left( n_{m+1,-s} - \frac{1}{2} \right) 
  \nonumber \\
  & & - \frac{U^{3}}{4} \sum_{m=1}^{N} 
  \left( n_{m \uparrow} - \frac{1}{2} \right) 
  \left( n_{m \downarrow} - \frac{1}{2} \right),
  \label{eq.I3} \\
  I^{(4)} & = &{\rm i} \sum_{m=1}^{N} \sum_{s=\uparrow \downarrow} 
  \left( c_{m+4 s}^{\dagger} c_{m s} - c_{m s}^{\dagger} c_{m+4 s} 
  \right) \nonumber \\
  & & - 2 {\rm i}U \sum_{m=1}^{N} \sum_{s=\uparrow \downarrow} 
  \Bigg\{ \left( c_{m+3 s}^{\dagger} c_{m s} 
  - c_{m s}^{\dagger} c_{m+3 s} \right) \sum_{k=m}^{m+3} 
  \left( n_{k,-s} - \frac{1}{2} \right) \nonumber \\
  & & \ \ \ \ \ \ \ \ \ \ \ \ \ \ \ \ \ \ 
  - \left( c_{m+1 s}^{\dagger} c_{m s} 
  - c_{m s}^{\dagger} c_{m+1 s} \right) \sum_{k=m-1}^{m+2} 
  \left( n_{k,-s} - \frac{1}{2} \right) \nonumber \\   
  & & \ \ \ \ \ \ \ \ \ \ \ \ \ \ \ \ \ \ \ \ 
  + \left( c_{m+2 s}^{\dagger} c_{m s} 
  + c_{m s}^{\dagger} c_{m+2 s} \right) 
  \sum_{k=m-1}^{m+2} \left( c_{k+1,-s}^{\dagger} c_{k,-s} 
  - c_{k,-s}^{\dagger} c_{k+1,-s} \right) \Bigg\} \nonumber  \\
  & & + 4 {\rm i} U^2 \sum_{m=1}^{N} 
  \sum_{s = \uparrow \downarrow} 
  \Bigg\{ \left(c_{m+2 s}^{\dagger} c_{m s} 
  - c_{m s}^{\dagger} c_{m+2 s} \right) 
  \big\{ \left( n_{m,-s} - \frac{1}{2} \right) 
  \left( n_{m+1,-s} - \frac{1}{2} \right) \nonumber \\
  & & \ \ \ \ \ \ + \left( n_{m,-s} - \frac{1}{2} \right) 
  \left( n_{m+2,-s} - \frac{1}{2} \right) 
  + \left( n_{m+1,-s} - \frac{1}{2} \right) 
  \left( n_{m+2,-s} - \frac{1}{2} \right) \big\} \nonumber \\
  & & \ \ \ \ \left( c_{m+1 s}^{\dagger} c_{m s} 
  + c_{m s}^{\dagger} c_{m+1 s} \right) 
  \big\{ \left( c_{m-1,-s}^{\dagger} 
  c_{m,-s} - c_{m,-s}^{\dagger} c_{m+1,-s} \right) 
  \left( n_{m+1,-s} - \frac{1}{2} \right) \nonumber \\
  & & \ \ \ \ \ \ \ \ \ \ 
  + \left( c_{m+1,-s}^{\dagger} c_{m+2,-s} 
  - c_{m+2,-s}^{\dagger} c_{m+1,-s} \right) 
  \left( n_{m,-s} - \frac{1}{2} \right) \big\} \Bigg\} 
  \nonumber \\
  & & + 2 {\rm i} U^{3} 
  \sum_{m=1}^{N} \sum_{s=\uparrow \downarrow} 
  \left( c_{m+1 s}^{\dagger} c_{m s} 
  - c_{m s}^{\dagger} c_{m+1 s} \right) 
  \left( n_{m+1,-s} + n_{m,-s} -1 \right). \label{eq.I4}
\end{eqnarray} 
These currents are embedded in the fermionic transfer matrix 
(\ref{eq.transfermatrix}) and should be ${SO(4)}$ 
invariant from the result in the previous section. 
One can confirm the ${SO(4)}$ invariance of 
these currents using the transformation law of the fermion 
operators (\ref{eq.summaryfermion}). 
In the following, we present a different approach; we shall rewrite 
these currents in manifestly ${SO(4)}$ invariant forms
by use of the Clifford algebra. 

Define  ${\G_{j}^{a} \ (j=1, \cdots, N, a=1,\cdots, 4)}$ by
\begin{eqnarray}
  & & \G_{2n-1}^{1} = c_{2n-1 \uparrow}^{\dagger} 
  + c_{2n-1 \uparrow}, \ \ 
  \G_{2n-1}^{2} = {\rm i} \left( c_{2n-1 \uparrow}^{\dagger} 
  - c_{2n-1 \uparrow} \right), \nonumber \\
  & & \G_{2n-1}^{3} = c_{2n-1 \downarrow}^{\dagger} 
  + c_{2n-1 \downarrow}, \ \ 
  \G_{2n-1}^{4} = {\rm i} 
  \left( c_{2n-1 \downarrow}^{\dagger} 
  - c_{2n-1 \downarrow} \right),
\end{eqnarray}
and
\begin{eqnarray}
  & & \G_{2n}^{1} = {\rm i} 
  \left( c_{2n \uparrow} - c_{2n \uparrow}^{\dagger} \right), \ \ 
  \G_{2n}^{2} = c_{2n \uparrow}^{\dagger} + c_{2n \uparrow}, 
  \nonumber \\
  & & \G_{2n}^{3} = {\rm i} 
  \left( c_{2n \downarrow} - c_{2n \downarrow}^{\dagger} \right), \ \ 
  \G_{2n}^{4} =  c_{2n \downarrow}^{\dagger} + c_{2n \downarrow},
\end{eqnarray}
where ${n=1,\cdots,\frac{N}{2}}$.
Then the operators ${\G_{j}^{a}}$ satisfy 
the defining relations of the Clifford algebra \cite{Affleck,Noga}
\begin{equation}
\ \ \ \ \ \ \ \ \ \ \ \ \ \ \left\{ \G_{j}^{a}, \G_{k}^{b} \right\} 
  = 2 \delta_{jk} \delta^{a b}, \ \ \ \ j,k = 1, \cdots, N, 
\ a,b = 1,\cdots, 4. \label{eq.Clifford}
\end{equation}
In terms of ${\G_{j}^{a}}$, 
 the ${SO(4)}$ rotation for the fermion operators (\ref{eq.summaryfermion}) 
can be expressed simply as
\begin{equation}
  \tilde{\G}_{j}^{a} = \sum_{b=1}^{4} G^{a b} \G_{j}^{b}, \ \ \ \ \ 
  G = (G^{a b}) \in SO(4).  
  \label{eq.GammaSO4}
\end{equation}
The relation between the matrices ${G}$ and ${M}$ in \S 4 is 
explicitly given by 
\begin{eqnarray*}
&&  G = G^{(1)} G^{(2)}, \nonumber \\ 
&&  G^{(1)} = \left( 
  \begin{array}{cccc}
    \xi_{0}  & \xi_{1} &   \xi_{3} & - \xi_{2} \\
  - \xi_{1}  & \xi_{0} & - \xi_{2} & - \xi_{3} \\  
  - \xi_{3}  & \xi_{2} &   \xi_{0} &   \xi_{1} \\
    \xi_{2}  & \xi_{3} & - \xi_{1} &   \xi_{0}
  \end{array}
  \right), \ \ G^{(2)} = \left( 
  \begin{array}{cccc}
    \zeta_{0}  &   \zeta_{1} & - \zeta_{3} & - \zeta_{2} \\
  - \zeta_{1}  &   \zeta_{0} &   \zeta_{2} & - \zeta_{3} \\
    \zeta_{3}  & - \zeta_{2} &   \zeta_{0} & - \zeta_{1} \\
    \zeta_{2}  &   \zeta_{3} &   \zeta_{1} &   \zeta_{0}
  \end{array}
  \right), 
\end{eqnarray*}
where ${\xi_{i}}$ and ${\eta_{i}}$ are real numbers given by
\begin{eqnarray}
  & & \xi_{0} = {\rm Re}(M_{11}), \ \ \ \ 
      \xi_{1} = {\rm Im}(M_{11}), \ \ \ \ 
      \xi_{2} = {\rm Re}(M_{14}), \ \ \ \ 
      \xi_{3} = {\rm Im}(M_{14}), \nonumber \\
  & & \zeta_{0} = {\rm Re}(M_{22}), \ \ \ \ 
      \zeta_{1} = {\rm Im}(M_{22}), \ \ \ \ 
      \zeta_{2} = {\rm Re}(M_{23}), \ \ \ \ 
      \zeta_{3} = {\rm Im}(M_{23}), \nonumber \\
 & &  \ \ \ \ \ \ \ \ \ \ \ \ \ \ \ \ \ \ \ \ \ 
      \ \ \ \ \ \ \ \ \ \ \ \ \ \ 
 \sum_{j=0}^{3} \xi_{j}^2 = \sum_{j=0}^{3} \zeta_{j}^{2} = 1.
\end{eqnarray}
Clearly the matrix ${G^{(1)}}$ corresponds to 
the charge-${SU(2)}$ and the matrix ${G^{(2)}}$ 
corresponds to the spin-${SU(2)}$. 
It is an interesting exercise to confirm that the matrices 
${G^{(1)}}$ and ${G^{(2)}}$ commute each other
\[
  G^{(1)}G^{(2)} = G^{(2)}G^{(1)}.
\]

Define the operator ${\G_{j}^{5}}$ by
\begin{equation}
  \G_{j}^{5} = \G_{j}^{1} \ \G_{j}^{2} \ \G_{j}^{3} \ \G_{j}^{4} 
  =  \frac{1}{4!} \sum_{\ a,\cdots,d=1}^{4}
  \epsilon_{a b c d} \ \G_{j}^{a} \ \G_{j}^{b} \ \G_{j}^{c} \ \G_{j}^{d}.
  \label{eq.defGamma5}
\end{equation}
The operator ${\G_{j}^{5}}$ has the following properties
\begin{equation}
 \ \ \ \  \left\{ \G_{j}^{5}, \G_{j}^{a} \right\} = 0, \ \ 
 \ \ \left[ \G_{j}^{5}, \G_{k}^{a} \right] = 0 \ \ \ \ (j \ne k), \ \ 
 \ \ a=1, \cdots, 4. 
  \label{eq.Gamma5}
\end{equation}
It is clear that the operators such as
\begin{equation}
 \sum_{a=1}^{4} \G_{j}^{a} \ \G_{k}^{a}, 
 \ \ \ \ \G_{j}^{5}, \label{eq.SO4term}
\end{equation} 
are invariant under the ${SO(4)}$ rotation (\ref{eq.GammaSO4}).
We make use of this fact to rewrite the conserved currents.
The Hamiltonian ${{\cal H} = I^{(1)}}$ 
in terms of the operators ${\G_{j}^{a}}$ and ${\G_{j}^{5}}$ 
is \cite{Affleck,Noga}
\begin{equation}
  I^{(1)} =\sum_{j} \sum_{a} (-1)^{j} 
  \G_{j+1}^{a} \G_{j}^{a} + u \sum_{j} \G_{j}^{5}, \label{eq.GammaI1}
\end{equation}
where
\[
  u = \frac{{\rm i} U}{2}.
\]
Here and hereafter, we do not mind the difference of an overall factor. 
The formula (\ref{eq.GammaI1}) gives a manifestly ${SO(4)}$ invariant 
representation of the Hamiltonian.

In the same way, we express the higher conserved currents 
in terms of the operators ${\G_{j}^{a}}$ and ${\G_{j}^{5}}$ as
\begin{eqnarray}
  I^{(2)} & = & \sum_{j} \sum_{a} \G_{j+2}^{a} \G_{j}^{a} 
  + u \sum_{j} \sum_{a} (-1)^{j} \G_{j+1}^{a} \G_{j}^{a} 
  \left( \G_{j+1}^{5} - \G_{j}^{5} \right), \label{eq.GammaI2} \\
  I^{(3)} & = & \sum_{j} \sum_{a} (-1)^{j} 
  \G_{j+3}^{a} \G_{j}^{a} \nonumber \\
  & & - u \sum_{j} \Bigg\{ \G_{j}^{5} 
  + \sum_{a} \G_{j+2}^{a} \G_{j}^{a} 
  \left( \G_{j+2}^{5} - \G_{j}^{5} \right) \nonumber \\
  & & \ \ \ \ \ \ \ \ \ \ \
  + \sum_{a \ne b} 
  \left( \G_{j+2}^{a}\G_{j+1}^{a} \G_{j+1}^{b}  
  \G_{j}^{b} \G_{j+1}^{5} - \frac{1}{2} \G_{j+1}^{a} 
  \G_{j+1}^{b} \G_{j}^{a} \G_{j}^{b} \G_{j}^{5} \right) \Bigg\} 
  \nonumber \\
  & & + u^2 \sum_{j} \sum_{a} (-1)^j \G_{j+1}^{a} \G_{j}^{a} 
  \G_{j+1}^{5} \G_{j}^{5} \nonumber \\
  & & + u^3 \sum_{j} \G_{j}^{5}, \label{eq.GammaI3} \\ 
  I^{(4)} & = & \sum_{j} \sum_{a} \G_{j+4}^{a} \G_{j}^{a} \nonumber \\
  & & + u \sum_{j} (-1)^{j} \Bigg\{ \sum_{a}  
  \left\{ \G_{j+3}^{a} \G_{j}^{a} 
  \left( \G_{j+3}^{5} - \G_{j}^{5} \right) 
  - \G_{j+1}^{a} \G_{j}^{a} 
  \left( \G_{j+1}^{5} - \G_{j}^{5} \right) \right\} 
  \nonumber \\
  & & \ \ \ \ \ \ \ \ \ \ \ \ \ \ \ \ \ \   
  + \sum_{a \ne b} \Big\{ \G_{j+3}^{a} \G_{j+2}^{a} 
  \G_{j+2}^{b} \G_{j}^{b} \G_{j+2}^{5} 
  + \G_{j+3}^{a} \G_{j+1}^{a} \G_{j+1}^{b} 
  \G_{j}^{b} \G_{j+1}^{5} \nonumber \\
  & & \ \ \ \ \ \ \ \ \ \ \ \ \ \ \  
      \ \ \ \ \ \ \ \ \ \ \ \ \ \ \ \ 
  - \G_{j+2}^{a} \G_{j+2}^{b} \G_{j+1}^{a} 
  \G_{j}^{b} \G_{j+2}^{5}  + \G_{j+2}^{a} 
  \G_{j+1}^{b} \G_{j}^{a} \G_{j}^{b} \G_{j}^{5} 
  \Big\} \Bigg\} \nonumber \\
  & & + u^2 \sum_{j} \Bigg\{ 
  \sum_{a} \G_{j+2}^{a} \G_{j}^{a} \G_{j+2}^{5} \G_{j}^{5} 
  - \sum_{a \ne b} \G_{j+2}^{a} \G_{j+1}^{a} 
  \G_{j+1}^{b} \G_{j}^{b} \G_{j+1}^{5}
  \left( \G_{j+2}^{5} - \G_{j}^{5} \right) \Bigg\} 
  \nonumber \\
  & & + u^3 \sum_{j} \sum_{a} (-1)^{j} 
  \G_{j+1}^{a} \G_{j}^{a} 
  \left( \G_{j+1}^{5} - \G_{j}^{5} \right). \label{eq.GammaI4}
\end{eqnarray}
Since the terms that constitute (\ref{eq.GammaI2})--(\ref{eq.GammaI4}) 
are of the form (\ref{eq.SO4term}), we can see that 
the higher conserved currents ${I^{(2)},I^{(3)}}$ and ${I^{(4)}}$ 
are also manifestly ${SO(4)}$ invariant. 
Note that the constraints ${a \ne b}$ in the summations
do not break the ${SO(4)}$ symmetry. For example, we can write
\begin{equation}
  \sum_{a \ne b} \G_{j+2}^{a} \G_{j+1}^{a} 
  \G_{j+1}^{b} \G_{j}^{b} \G_{j+1}^{5}
  = \sum_{a,b} \G_{j+2}^{a} \G_{j+1}^{a} \G_{j+1}^{b} 
  \G_{j}^{b} \G_{j+1}^{5}
  - \sum_{a} \G_{j+2}^{a} \G_{j}^{a} \G_{j+1}^{5}. 
  \label{eq.partialSO4}
\end{equation}
Both terms in the RHS of (\ref{eq.partialSO4}) are 
clearly ${SO(4)}$ invariant.

The infinitesimal generators of the ${SO(4)}$ rotations 
(\ref{eq.GammaSO4}) are given by \cite{Georgi}
\begin{equation}
  Q^{ab} = - Q^{ba} = \frac{1}{4 {\rm i}} \sum_{j}  
  \left[ \G_{j}^{a}, \G_{j}^{b} \right].  \label{eq.so4generator}
\end{equation}
In fact the generator (\ref{eq.so4generator}) fulfills %
the defining relation of the Lie algebra ${so(4)}$
\begin{equation}
  \left[ Q^{ab}, Q^{cd} \right] = 
  -{\rm i} \left( \delta^{bc} Q^{ad} - \delta^{ac} Q^{bd} 
  - \delta^{bd} Q^{ac} + \delta^{ad} Q^{bc} \right).
\end{equation}
 We also have a relation
\begin{equation}
\left[ Q^{ab}, \G_{j}^{c} \right] = {\rm i} \left( \delta^{ac} \G_{j}^{b} %
- \delta^{bc} \G_{j}^{a} \right), \label{eq.commuQandGamma}
\end{equation}
which is nothing but the infinitesimal %
transformation of (\ref{eq.GammaSO4}).
Using (\ref{eq.commuQandGamma}), one can confirm the commutativity
\begin{equation}
  \left[ Q^{ab}, I^{(n)} \right] = 0, \ \ \ \ \ \  n=1, \cdots,4,
\end{equation}
which shows the Lie algebra ${so(4)}$ symmetry 
of the conserved currents ${I^{(n)}}$.
Actually the generators of the spin-${su(2)}$ (\ref{eq.spinsu2}) 
and the charge-${su(2)}$ (\ref{eq.etasu2}) are related to ${Q^{ab}}$ as
\begin{eqnarray}
S^{x}     &=& -\frac{1}{2} \left( Q^{14} - Q^{23} \right), 
\ \ S^{y}= \frac{1}{2} \left( Q^{24} + Q^{13} \right), 
\ \ S^{z} = -\frac{1}{2} \left( Q^{12} - Q^{34} \right), 
\label{eq.spinGamma} \\ 
\eta^{x}  &=& -\frac{1}{2} \left( Q^{14} + Q^{23} \right), 
\ \ \eta^{y} = \frac{1}{2} \left( Q^{24} - Q^{13} \right), 
\ \ \eta^{z} = -\frac{1}{2} \left( Q^{12} + Q^{34} \right), 
\label{eq.chargeGamma}
\end{eqnarray}
where we introduced ${S^{x},S^{y},\eta^{x}}$ and 
${\eta^{y}}$ through the relations
\[
S^{\pm} = S^{x} \pm {\rm i} S^{y}, \ \ %
\ \ \eta^{\pm} = \eta^{x} \pm {\rm i} \eta^{y}.
\]

 For the Clifford algebra (\ref{eq.Clifford}), 
the partial particle-hole transformation 
(\ref{eq.particlehole}) corresponds to 
\begin{equation}
\G_{j}^{3} \longrightarrow - \G_{j}^{3}. \label{eq.exchangeGamma3}
\end{equation}
 Note that the transformation (\ref {eq.exchangeGamma3}) 
 exchanges the spin-${su(2)}$ (\ref{eq.spinGamma}) and 
 the charge-${su(2)}$ (\ref{eq.chargeGamma}). 
 As for the conserved currents, the transformation 
 (\ref{eq.exchangeGamma3}) preserves the operators 
 like ${\displaystyle \sum_{a=1}^{4} \G_{j}^{a} \ \G_{k}^{a}}$, 
 but changes the sign of ${\G_{j}^{5}}$.
From the explicit formulas (\ref{eq.GammaI1})--(\ref{eq.GammaI4}), 
one can immediately find that the conserved currnets ${I^{(n)} (n =1,\cdots,4)}$ 
are invariant under the partial particle-hole transformation 
\begin{equation}
\G_{j}^{5} \longrightarrow - \G_{j}^{5}, \ \ \ \ u \longrightarrow -u.
\end{equation}
This is consistent with the result in \S 4. 
\vspace{20pt}
\begin{flushleft}
{\large \bf \S 6. Discussions}
\end{flushleft}
\setcounter{equation}{0}
\renewcommand{\theequation}{6.\arabic{equation}}
We have investigated the ${SO(4)}$ symmetry of 
the 1D Hubbard model from the QISM point of view. 
Our approach is based on the fermionic formulation of 
the Yang-Baxter relation for the 1D Hubbard model 
found by Olmedilla et al. \cite{Olmedilla1}.
It consists of the fermionic ${R}$-matrix and 
the fermionic ${L}$-operator. 
We have discovered the transformation law (\ref{eq.evenoddSO4}) of 
the fermionic ${L}$-operator under the ${SO(4)}$ rotation. 
It is  a kind of gauge transformation and induces 
the transformation of the monodromy matrix. 
The result is a fundamental property of the associative algebra 
defined by the fermionic ${R}$-matrix of the 1D Hubbard model. 
We believe that the property will also play an important role 
in the algebraic Bethe ansatz for the 1D Hubbard model, 
which was recently explored by Ramos and Martins \cite{Ramos}. 
We like to emphasize the advantage of the fermionic formulation 
of the Yang-Baxter relation. 
It is difficult to discuss the ${SO(4)}$ symmetry of 
the Hubbard model through Shastry's 
${R}$-matrix and the related transfer matrix. 

The ${SO(4)}$ invariance of the transfer matrix ensures
the ${SO(4)}$ invariance of the conserved currents. 
We have demonstrated the ${SO(4)}$ symmetry of 
the higher conserved currents employing the Clifford algebra,
which corresponds to the spinor representation of the rotation group. 
It should be interesting to explore a representation of 
the fermionic ${L}$-operator itself in terms of the Clifford algebra.
 
On the infinite lattice, the Lie algebra 
${so(4) = su(2) \oplus su(2)}$ symmetry of the 1D Hubbard model 
is extended to the Yangian 
${{\rm Y}(so(4)) = {\rm Y}(su(2)) \oplus {\rm Y}(su(2))}$ symmetry
\[
  \left[ {\rm Y}(so(4)), I^{(1)} \right] = 0
\]
as was discovered by Uglov and Korepin \cite{Uglov}. 
The generators of ${{\rm Y}(so(4))}$ can be expressed 
in terms of the Clifford algebra ${\G_{j}^{a}}$ as follows
\begin{eqnarray*}
  Q_{ab}^{(0)} &=& - \frac{\rm i}{4} \sum_{j} \left[ \G_{j}^{a},  
                   \G_{j}^{b} \right], \\
  Q_{ab}^{(1)} &=& - {\rm i} \sum_{j} (-1)^{j} 
  \left( \G_{j+1}^{a} \G_{j}^{b} 
  + \G_{j}^{a} \G_{j+1}^{b} \right) \nonumber \\
  & & + \frac{{\rm i} u}{4} 
  \left( \sum_{j > k} - \sum_{ k > j} \right) 
  \sum_{c \ne a, b} \G_{j}^{a} \G_{j}^{c} 
  \G_{k}^{c} \G_{k}^{b} 
  \left( \G_{j}^{5} + \G_{k}^{5} \right).
\end{eqnarray*}
By use of the fundamental properties of 
the Clifford algebra (\ref{eq.Clifford}), 
(\ref{eq.defGamma5}) and (\ref{eq.Gamma5}), 
we have confirmed that 
the higher conserved currents ${I^{(n)} (n=2, 3, 4)}$ 
also have the Yangian ${{\rm Y}(so(4))}$ symmetry, i.e.,
\[
  \left[ Q_{ab}^{(0)}, I^{(n)} \right] 
  = \left[ Q_{ab}^{(1)}, I^{(n)} \right] = 0, \ \ \ \ n = 1, \cdots, 4.
\]
All the conserved currents of the 1D Hubbard model 
on the infinite lattice are conjectured to have the 
${{\rm Y}(so(4))}$ symmetry. 
In fact Murakami and G${\ddot{\rm o}}$hmann \cite{Murakami}
recently showed the existence of 
an infinite number of the conserved currents 
which have the Yangian symmetry on the infinite lattice. 
However, one of the two ${{\rm Y}(su(2))}$ 
that constitute ${{\rm Y}(so(4))}$ drops out \cite{Murakami}. 
It seems to be difficult to prove the full ${{\rm Y}(so(4))}$ 
symmetry of the conserved currents simultaneously in their method. 

\vspace{20pt}
\begin{flushleft}   
{\large \bf Acknowledgements}
\end{flushleft}

The authors are grateful to F. G${\ddot{\rm o}}$hmann, K. Hikami, 
S. Murakami and T. Tsuchida for valuable discussions and comments.
One of the authors (M. S.) also thank J. Suzuki and A. Kuniba for
discussions. This work is supported by a Grant-in-Aid for JSPS Fellows
from the Ministry of Education, Science, Sports and Culture of Japan.

\vspace{20pt}

\begin{flushleft}
{\large \bf Appendix : Symmetry Matrix of Shastry's {\mbox{\boldmath $R$}}-Matrix}
\end{flushleft}
\setcounter{equation}{0}
\renewcommand{\theequation}{A.\arabic{equation}}
The ${L}$-operator for the coupled spin model (\ref{eq.Coupled}) %
\cite{Shastry1,Shastry2,Shastry3,Olmedilla1} is expressed as
\begin{equation}
L_{m}(\theta) = \left(
              \begin{array}{cccc}
   {\rm e}^{h} p_{m}^{+}(\theta) q_{m}^{+}(\theta) %
   & p_{m}^{+}(\theta) \tau_{m}^{-} %
   & \sigma_{m}^{-} q_{m}^{+}(\theta) %
   & {\rm e}^{h} \sigma_{m}^{-} \tau_{m}^{-} \\
   p_{m}^{+}(\theta) \tau_{j}^{+} %
   & {\rm e}^{-h} p_{m}^{+}(\theta) q_{m}^{-}(\theta) %
   & {\rm e}^{-h} \sigma_{m}^{-} \tau_{m}^{+} %
   & \sigma_{m}^{-} q_{m}^{-}(\theta) \\
   \sigma_{m}^{+} q_{m}^{+}(\theta) %
   & {\rm e}^{-h} \sigma_{m}^{+} \tau_{m}^{-} %
   & {\rm e}^{-h} p_{m}^{-}(\theta) q_{m}^{+}(\theta) %
   & p_{m}^{-}(\theta) \tau_{m}^{-} \\
   {\rm e}^{h} \sigma_{m}^{+} \tau_{m}^{+} %
   & \sigma_{m}^{+} q_{m}^{-}(\theta) %
   & p_{m}^{-}(\theta) \tau_{m}^{+} %
   & {\rm e}^{h} p_{m}^{-}(\theta) q_{m}^{-}(\theta)
              \end{array}
               \right), \label{eq.ShastryL}
\end{equation}
where
\begin{eqnarray}
p_{m}^{\pm}(\theta) & = & \frac{1}{2} 
\left( \cos \theta  + \sin \theta  \right) \pm \frac{1}{2} 
\left( \cos \theta - \sin \theta \right) \sigma_{m}^{z}, \nonumber \\
q_{m}^{\pm}(\theta) & = & \frac{1}{2} 
\left( \cos \theta + \sin \theta  \right) \pm \frac{1}{2} 
\left( \cos \theta - \sin \theta  \right) \tau_{m}^{z}.
\end{eqnarray}
Here the coupling constant ${h}$ is related to the spectral %
parameter ${\theta}$ through the formula (\ref{eq.constraint1}).
The ${R}$-matrix ${\check{R}_{12}(\theta_1,\theta_2)}$, 
which satisfies the Yang-Baxter relation (\ref{eq.ShastryYBR}) 
with the ${L}$-operator (\ref{eq.ShastryL}),
is connected to the fermionic ${R}$-matrix (\ref{eq.Rmatrix}) through 
the formula \cite{Olmedilla1}
\begin{equation}
\check{R}_{12}(\theta_1,\theta_2) = 
W_{12}^{-1} \check{\cR}_{12}(\theta_1,\theta_2) W_{12},
\end{equation}
where ${W_{12}}$ is a diagonal ${16 \times 16}$ matrix
\begin{equation}
W_{12} = \rm{diag}(1,1,- \rm{i},- \rm{i},- \rm{i},
- \rm{i},1,1,-1,-1,\rm{i},\rm{i},\rm{i},\rm{i},-1,-1).
\end{equation}

We consider the symmetry matrix ${M}$ of 
Shastry's ${R}$-matrix ${\check{R}_{12}(\theta_1,\theta_2)}$ 
which is defined to be a constant matrix satisfying 
\begin{equation}
\left[ \check{R}_{12}(\theta_1,\theta_2), M \otimes M \right] = 0. 
\label{eq.defsymmetryShastry}
\end{equation}
Here the matrix elements of ${M}$ are assumed to be commuting numbers. %
One might suppose the symmetry matrix of Shastry's ${R}$-matrix 
is identical to that of the fermionic ${R}$-matrix 
(\ref{eq.symmetrymatrix}). However, surprisingly enough, we notice 
they take different forms. In fact, solving the defining equation 
for the symmetry matrix (\ref{eq.defsymmetryShastry}) directly, 
we find that the followings are the symmetry matrix 
of Shastry's $R$-matrix,
\begin{equation}
M = \left( 
  \begin{array}{cccc}
    \alpha  &  0 &  0 &  0      \\
    0  & \beta &  0 &  0      \\
    0  &  0 & \gamma & 0      \\
    0  &  0 & 0 & \delta 
  \end{array}
  \right) , \ \ 
  \left( 
  \begin{array}{cccc}
    \alpha  &  0 &  0 &  0      \\
    0  &  0 &  \beta &  0      \\
    0  & \gamma & 0 & 0      \\
    0  &  0 & 0 & \delta 
  \end{array}
  \right) , \ \ \left( 
   \begin{array}{cccc}
    0 &  0 &  0 & \alpha      \\
    0  & \beta &  0 &  0      \\
    0  &  0 & \gamma & 0      \\
    \delta &  0 & 0 & 0 
  \end{array}
  \right) , \ \ 
  \left( 
  \begin{array}{cccc}
    0 &  0 &  0 & \alpha      \\
    0  &  0 & \beta &  0      \\
    0  & \gamma & 0 & 0      \\
   \delta  &  0 & 0 & 0 
  \end{array}
  \right) , \label{eq.SymmetryShastry}
\end{equation}
where ${\alpha, \beta, \gamma}$ and  
${\delta}$ are ${c}$-numbers obeying
\begin{equation}
\alpha \delta = \beta \gamma. \label{eq.SymmetryShastryCond} 
\end{equation}
We ignore the difference of overall factors 
of the matrices.
Then each matrix (\ref{eq.SymmetryShastry}) depends only on 
two parameters. 
The result means that Shastry's ${R}$-matrix does not reflect 
the ${SO(4)}$ symmetry of the fermionic Hamiltonian (\ref{eq.Hubbard}). 
Therefore the ${SO(4)}$ symmetry of the transfer matrix 
that we explored in this paper may not be discussed
if we use Shastry's ${R}$-matrix and ${L}$-operator 
(\ref{eq.ShastryL}).

\end{document}